# The Spin-Orbit Torque from a Magnetic Heterostructure with High-Entropy Alloy

Tian-Yue Chen[1], Tsao-Chi Chuang[2], Ssu-Yen Huang[2], Hung-Wei Yen[1], and Chi-Feng Pai[1]*

[1]*Department of Materials Science and Engineering, National Taiwan University, Taipei 10617, Taiwan*

[2]*Department of Physics, National Taiwan University, Taipei 10617, Taiwan*

High-entropy alloy (HEA) is a family of metallic materials with nearly equal partitions of five or more metals, which might possess mechanical and transport properties that are different from conventional binary or tertiary alloys. In this work, we demonstrate current-induced spin-orbit torque (SOT) magnetization switching in a Ta-Nb-Hf-Zr-Ti HEA-based magnetic heterostructure with perpendicular magnetic anisotropy (PMA). The maximum damping-like SOT efficiency from this particular HEA-based magnetic heterostructure is further determined to be $\left|\zeta_{DL}^{\text{HEA}}\right| \approx 0.033$ by hysteresis loop shift measurements, while that for the Ta control sample is $\left|\zeta_{DL}^{\text{Ta}}\right| \approx 0.04$. Our results indicate that HEA-based magnetic heterostructures can serve as a new group of potential candidates for SOT device applications.

---

* Email: cfpai@ntu.edu.tw

In recent years, high-entropy alloys (HEAs) have attracted considerable attention from the materials science community, since HEAs offer a new route of metallurgy research beyond traditional binary and/or tertiary alloy systems. HEAs typically consist of five or more principle elements with equimolar or nearly-equimolar ratios and can be viewed as composites at atomic scale [1,2], therefore the high mixing entropy. The mechanical properties of HEAs have been widely-studied, while their physical properties, especially spin-transport properties, have yet to be reported. Meanwhile, in the field of spintronics or spin-orbitronics, researchers have identified that the spin Hall effects (SHEs) [3-5] are strong in pure heavy transition metals (TM) such as Pt [6-8], β-Ta [9], and β-W [10] as well as in metallic alloys such as Cu-Bi [11], Cu-Ir [12], Cu-Pb [13], Au-W [14], Au-Cu [15], Pt-Hf (Al) [16], and even in antiferromagnetic alloying systems [17,18]. In order to extend the horizon of spin-Hall material exploration to beyond binary alloying systems, it would be very interesting and important to study the SHE in HEAs. In this work, we demonstrate the SHE-induced spin-orbit torque (SOT) magnetization switching from an HEA/Ta/CoFeB/Hf/MgO magnetic heterostructure, with the HEA being nominally-equimolar Ta-Nb-Hf-Zr-Ti. By using SOT-assisted hysteresis loop shift measurements [19], we find that the maximum damping-like SOT efficiency from our HEA-based magnetic heterostructures to be $\left|\varsigma_{DL}^{\text{HEA}}\right| \approx 0.033$, which is comparable to that from the Ta control samples $\left|\varsigma_{DL}^{\text{Ta}}\right| \approx 0.04$.

The HEA thin film (buffer layer) that we employed in our magnetic heterostructure was deposited from an equimolar $Ta_{20}Nb_{20}Hf_{20}Zr_{20}Ti_{20}$ (in at. %) sputter target. We picked this particular composition of HEA for several reasons. First of all, from the perspective of gaining interfacial perpendicular magnetic anisotropy (PMA) in a TM/CoFeB/MgO heterostructure, both Ta and Hf [20-23] are widely-used TM buffer layers for achieving this goal. Moreover, Zr has been shown to be a good boron sink to facilitate the crystallization of CoFe from amorphous CoFeB hence its templating with respect to the MgO (100)-orientation [24]. Recently, it has also been demonstrated that PMA can be obtained from both Zr/CoFeB/MgO and Nb/CoFeB/MgO systems after suitable heat treatments [25]. Secondly, a pure Ta in its highly resistive phase has been shown to be a strong spin Hall source [9,26], which can be utilized to efficiently generate SOTs. We would like to observe the change in terms of SOT efficiency if the Ta component has been diluted in a solid solution fashion. Lastly, though studies on the electron transport properties of HEAs are still rare, it has been reported that the $Ta_{34}Nb_{33}Hf_8Zr_{14}Ti_{11}$ HEA is a BCS type superconductor with a transition temperature $T_c \approx 7.3\,\text{K}$ [27]. Although the present work focuses on room-temperature transport properties of an HEA-based system with a different molar composition, the rich physics in high-entropy alloys make this particular HEA-based system attractive.

We prepared our heterostructure samples by DC (for metallic materials) and RF (for MgO) magnetron sputtering in a high vacuum system with a base pressure $\sim 3\times 10^{-8}$ Torr. The working

Ar pressure and the DC sputtering power for metallic materials deposition are 3 mTorr and 30 W, respectively. The condition for RF sputtering of MgO is 10 mTorr Ar with 50W forward power. The growth rates of each material were further characterized by X-ray reflectivity (XRR) and atomic force microscopy (AFM). To check the quality of our sputter-deposited Ta-Nb-Hf-Zr-Ti HEA film, we first performed cross-sectional high resolution transmission electron microscopy (HR-TEM, FEI Tecnai G2 F20) and in-situ X-ray energy dispersive spectrometry (EDS) on a 57 nm-thick HEA film, which was deposited on a Si/SiO$_2$ substrate. The TEM sample was prepared by a lift-out technique with Helios NanoLab 600i focus ion beam. As shown in Fig. 1(a), the HR-TEM image indicates that the sputtered HEA film is amorphous. The electrical resistivity of the HEA film is measured to be around $\rho_{\text{HEA}} \approx 138$ μΩ-cm, which is comparable to the reported values of other amorphous films that give rise to large spin Hall effects [26]. The in-situ EDS result (Fig. 1(b)) further indicates that the sputtered film composition to be fairly uniform across the whole thickness range, with its averaged atomic ratio being Ta$_{24.9}$Nb$_{18.7}$Hf$_{17.7}$Zr$_{18.3}$Ti$_{20.4}$. For simplicity, we will refer to the sputtered Ta$_{24.9}$Nb$_{18.7}$Hf$_{17.7}$Zr$_{18.3}$Ti$_{20.4}$ (nominal composition: Ta$_{20}$Nb$_{20}$Hf$_{20}$Zr$_{20}$Ti$_{20}$) film as "HEA" for the rest of the paper.

We utilized HEA-based PMA devices to investigate the SOT therein. For comparison, magnetic heterostructures with PMA were deposited on HEA and Ta buffer layers, as two series of samples: (I) HEA series: || HEA(3.5) /Ta(0.5)/Co$_{20}$Fe$_{60}$B$_{20}$($t_{\text{CoFeB}}$)/Hf(0.5)/MgO(2)/Ta(2) and

(II) Ta series (control samples): || Ta(4)/Co$_{20}$Fe$_{60}$B$_{20}$($t_{CoFeB}$)/Hf(0.5)/MgO(2)/Ta(2) (number in parentheses is film thickness in nanometers), where "||" stands for Si/SiO$_2$ substrate and $1.0\,\text{nm} \leq t_{CoFeB} \leq 1.6\,\text{nm}$ (nominal thickness). We further performed annealing for all samples at 300°C for 1 hour to promote PMA in the same high vacuum chamber after thin film depositions. The purpose of depositing Ta (0.5 nm) and/or Hf (0.5 nm) insertion layers on the opposite sides of CoFeB in these films is to facilitate the enhancement of PMA in these magnetic heterostructures [20,28]. Both insertions layers should have minimal effects on the spin-Hall transport properties of the whole structure, since the Ta layer thickness is ultra-thin and the Hf layer will be mostly oxidized by the MgO on-top [28].

The magnetic properties of both series of samples were characterized by vibrating sample magnetometer (VSM). As shown in Fig. 2 (a), the effective saturation magnetization of CoFeB in HEA series is $M_s = 1490\,\text{emu/cm}^3$ with a magnetic dead layer of $t_{dead} = 0.71\,\text{nm}$, while that for the Ta control series is $M_s = 1282\,\text{emu/cm}^3$ with $t_{dead} = 0.57\,\text{nm}$. The effective anisotropy energy densities (in terms of $K_{eff} \cdot t_{CoFeB}^{eff}$, where $t_{CoFeB}^{eff} = t_{CoFeB} - t_{dead}$) for both series of samples are shown in Fig. 2 (b). The interfacial contribution of anisotropy energy density, $K_s = 0.32\,\text{erg/cm}^2$ for the HEA series, was obtained by the intercept of extrapolating the linear regime of $K_{eff} \cdot t_{CoFeB}^{eff}$ vs. $t_{CoFeB}^{eff}$ plot [29]. Both series of samples show maximum perpendicular anisotropy at $t_{CoFeB}^{eff} \approx 0.7\,\text{nm}$. The overall magnetic property results indicate that sandwiching

CoFeB in between Ta and Hf dusting layers will give rise to similar $M_s$, $t_{dead}$, and $K_{eff}$, independent of the adopted buffer layer (either HEA or Ta). This feature allows us to study spin-transport properties that vary with the buffer layer or the spin-Hall source material, while keeping the magnetic layer or the spin current sink material (CoFeB) unchanged.

As schematically shown in Fig. 3 (a), we patterned samples into micron-sized Hall-bar devices with lateral dimensions of 5 μm by 60 μm for anomalous Hall (AH) voltage measurements. A representative AH loop was obtained from an HEA($t_{HEA}$)/Ta(0.5)/CoFeB(1.4)/Hf(0.5)/MgO(2)/Ta(2) sample with $t_{HEA} = 5$ nm, as shown in Fig. 3 (b), which suggests PMA indeed exists on the HEA buffer layer with out-of-plane coercive field $H_c \approx 10$ Oe. To demonstrate the efficacy of SOT from these HEA-based magnetic heterostructures, we performed current-induced SOT switching measurements. The current-induced switching data from this representative $t_{HEA} = 5$ nm Hall-bar device with opposite in-plane bias fields ($H_x = \pm 100$ Oe) are shown in Fig. 3 (c) and (d). The steps in current-induced switching data indicate that the switching is a domain nucleation/propagation process, which is similar to the switching curves in several previous reports [30,31]. The current-induced switching symmetry is also consistent with that from Ta-based [9] and W-based heterostructures [22,30], which indicates a negative spin Hall ratio for this particular HEA.

To further quantify SOT efficiencies of these HEA-based magnetic heterostructures,

especially the damping-like (Slonczewski-like) torque contribution, we performed AH voltage hysteresis loop shift measurements [19] on a series of HEA-based devices with various HEA thicknesses. The nominal CoFeB thickness in these heterostructures was fixed at 1.4 nm, since this particular thickness provides maximum $K_{eff}$. In Fig. 4 (a), we show the hysteresis loop shifts of an HEA ($t_{HEA} = 5$ nm) sample under the application of either a positive or a negative DC current $I_{DC}$ while simultaneously applying an in-plane bias field $H_x = 500\,\text{Oe}$. According to a SOT plus Dzyaloshinskii-Moriya interaction (SOT+DMI) scenario [19,32,33], the in-plane bias field $H_x$ will re-align Néel domain walls in the magnetic heterostructure while the DC current $I_{DC}$ flowing through the buffer layer (HEA in this case) will further generate SOT acting upon the re-aligned walls, thereby resulting in an out-of-plane effective field $H_{eff}$ that shifts the whole hysteresis loop. As shown in Fig. 4 (b), the linear trend between $H_{eff}$ and $I_{DC}$ then can be obtained to further calculate "current-induced effective field per current density" $\chi \equiv H_{eff} / J_e$, where $J_e$ is the charge current density flowing in the buffer layer. Typically, $\chi$ can be considered as the figure of merit for evaluating the SOT efficacy from different magnetic heterostructures. We summarize the in-plane field $H_x$ dependence of $\chi$ in Fig. 4 (c). When $H_x$ is large enough to fully realign the domain walls ($H_x = H_{sat} \approx 100$ Oe), the measured $\chi$ saturates at $\chi_{sat} \approx -17.4\,\text{Oe}/10^{11}\text{A/m}^2$. The magnitude of $\chi_{sat}$ is smaller yet comparable to reported values for pure Ta-based magnetic heterostructures [19,33]. The negative sign of $\chi_{sat}$ also confirms a negative spin Hall ratio (or spin Hall angle) of HEA, similar to that of Ta [9] and W [10] systems.

Based on the theory of SHE-induced SOT, $\chi_{sat}$ of a magnetic heterostructure with PMA can be related to its damping-like torque efficiency $\zeta_{DL}$ by $\chi_{sat} = (\pi/2)(\hbar\zeta_{DL}/2e\mu_0 M_s t_{FM}^{eff})$ [34], where $t_{FM}^{eff}$ is the effective thickness of ferromagnetic layer. Therefore, by using VSM-determined $M_s = 1490\,\text{emu/cm}^3$ ($1.49\times10^6\,\text{A/m}$ in SI units) and $t_{CoFeB}^{eff} \approx 0.69\,\text{nm}$, we can estimate $\zeta_{DL}$ of these samples. We summarize the magnitude of damping-like SOT efficiencies $|\zeta_{DL}|$ of HEA-based magnetic heterostructures with different HEA thicknesses ($t_{HEA}$) in Fig. 4 (d). The estimated $|\zeta_{DL}|$ for the HEA-based samples increases with respect to the buffer layer thickness, with a maximum of $|\zeta_{DL}^{HEA}| = 0.033 \pm 0.004$ at $t_{HEA} = 5\,\text{nm}$. This thickness dependence of $|\zeta_{DL}^{HEA}|$ can be well-fitted to a simple spin-diffusion model [6],

$|\zeta_{DL}(t_{HEA})| = |\zeta_{DL}(t_{HEA} \to \infty)| \cdot [1 - \text{sech}(t_{HEA}/\lambda_s^{HEA})]$, with spin diffusion length $\lambda_s^{HEA} = 2.5\,\text{nm}$ and $|\zeta_{DL}(t_{HEA} \to \infty)| = 0.04$. The results from two control samples, Ta-based heterostructures with PMA (Ta($t_{Ta} = 3.5\,\text{nm}$ and $4\,\text{nm}$)/CoFeB(1.4)/Hf(0.5)/MgO(2)/Ta(2)), are also plot in Fig. 4 (d) for comparison. It can be seen that the HEA-based magnetic heterostructures possess damping-like torque efficiencies that are about 50% to 80% of their Ta-based counterparts, though the composition of the major spin Hall material, Ta, is only 24.9% (at %) in the deposited HEA films. Only if we take the maximum possible Hf contribution (17.7%, with the largest reported $\left|\zeta_{DL}^{\text{Hf}}\right| \approx 0.11$ [26]) into account, the damping-like torque efficiency will reach $\left|\zeta_{DL}^{\text{HEA}}\right| \approx 0.03$ based on Vegard's mixing rule for solid solution. However, other reports also suggest that Hf has a limited spin Hall effect ($\left|\zeta_{DL}^{\text{Hf}}\right| \approx 0$) [22], especially when it's thin ($\leq 2$ nm) [35]. Therefore, we believe that our observation here cannot be simply explained by adopting Vegard's mixing rule. This deviation of the transport property from an ideal mixing scenario is consistent with the discovery that the electrical degree of freedom does not follow a "cocktail effect" of the constituent elements in an HEA [27], though the mixture is mostly random and in an amorphous solid solution fashion.

In conclusion, we show that Ta-Nb-Hf-Zr-Ti HEA-based magnetic heterostructures can possess PMA as well as SOT efficiencies that are comparable to those of Ta-based heterostructures, though the concentration of the major spin Hall or spin-orbit coupling sources, such as Ta and/or Hf, have been diluted down to less than 50% (at %). We demonstrate current-induced SOT switching in these HEA-based devices and further characterize the maximum damping-like SOT efficiency to be $\left|\zeta_{DL}^{\text{HEA}}\right| \approx 0.033$. Our discovery suggests that by working on randomly-mixing alloys beyond binary systems, it is possible to explore electronic degree of freedom that does not follow Vegard's mixing rule in other similar HEA-based magnetic heterostructures, thereby improving the SOT efficiency for possible next-generation SOT memory device applications.

**Acknowledgements**

This work was supported by the Ministry of Science and Technology of Taiwan under Grant No. MOST 105-2112-M-002-007-MY3 and No. MOST 103-2212-M-002-021-MY3. S. Y. Huang acknowledges the Golden Jade Fellowship of the Kenda Foundation, Taiwan. C. F. Pai would like to thank Yongxi Ou from Cornell University for fruitful discussions.

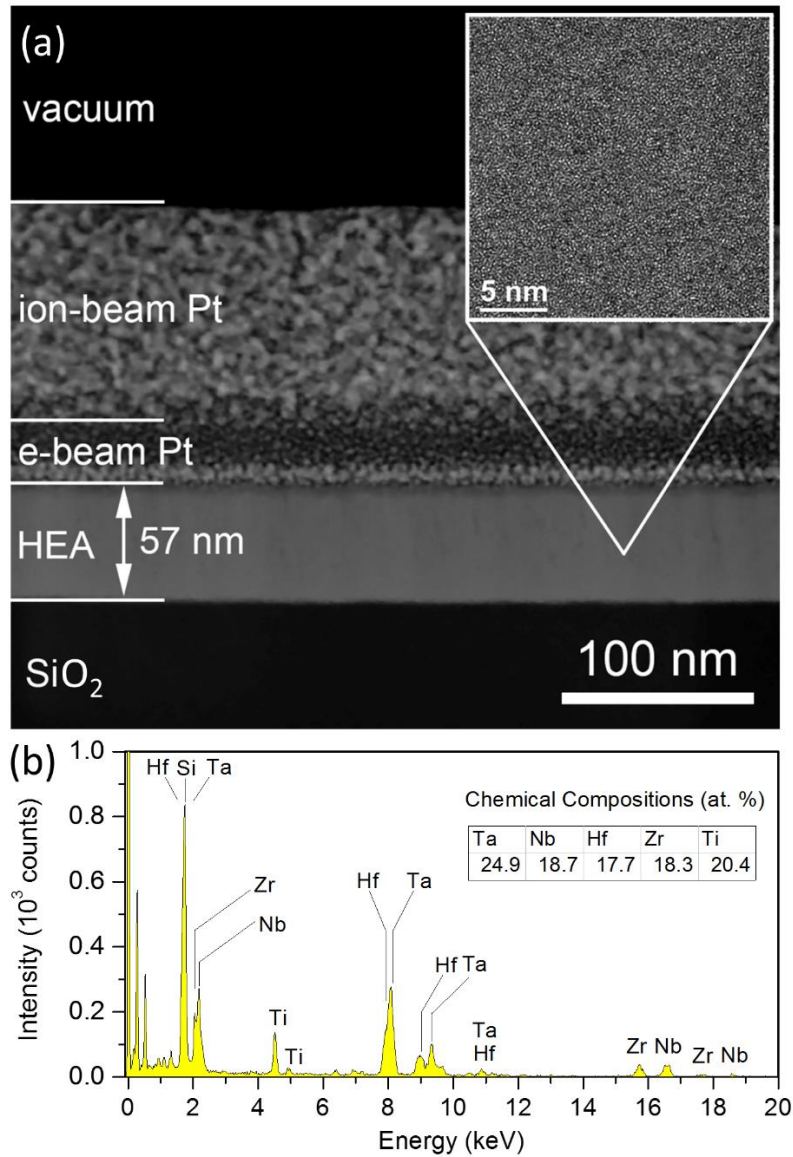

Figure 1. (a) Cross-sectional high resolution TEM (HR-TEM) image of a 57-nm thick Ta-Nb-Hf-Zr-Ti high-entropy alloy (HEA) film prepared by high vacuum sputter deposition. The Pt layers serve as protecting caps for the HEA underneath. (b) Energy-dispersive spectroscopy (EDS) result of the sputter-deposited Ta-Nb-Hf-Zr-Ti HEA film.

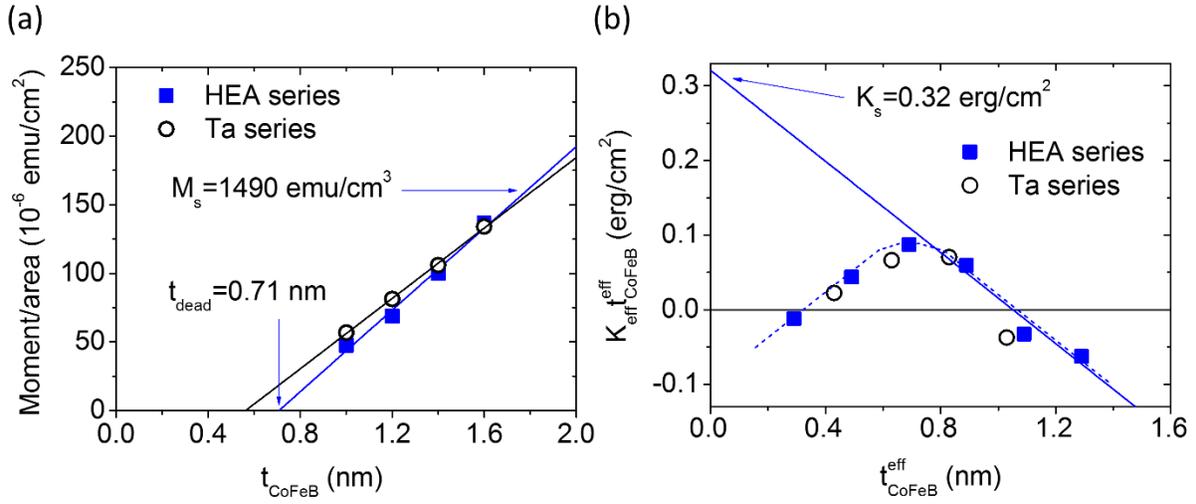

Figure 2. (a) Vibrating sample magnetometer (VSM) measurements of magnetizations in HEA series (blue squares): HEA(3.5)/Ta(0.5)/CoFeB($t_{CoFeB}$)/Hf(0.5)/MgO(2) and Ta series (black open circles): Ta(4)/CoFeB($t_{CoFeB}$)/Hf(0.5)/MgO(2)/Ta(2) samples. Solid lines represent linear fits to data. (b) VSM-determined effective magnetic anisotropy energy densities (in terms of $K_{eff} \cdot t_{CoFeB}^{eff}$) for HEA (blue squares) and Ta (black open circles) series samples. The blue solid line represents the linear fit of data with $t_{CoFeB}^{eff} = t_{CoFeB} - t_{dead} \geq 0.8\,\text{nm}$ for HEA series samples. The dashed line serves as guide to the eye.

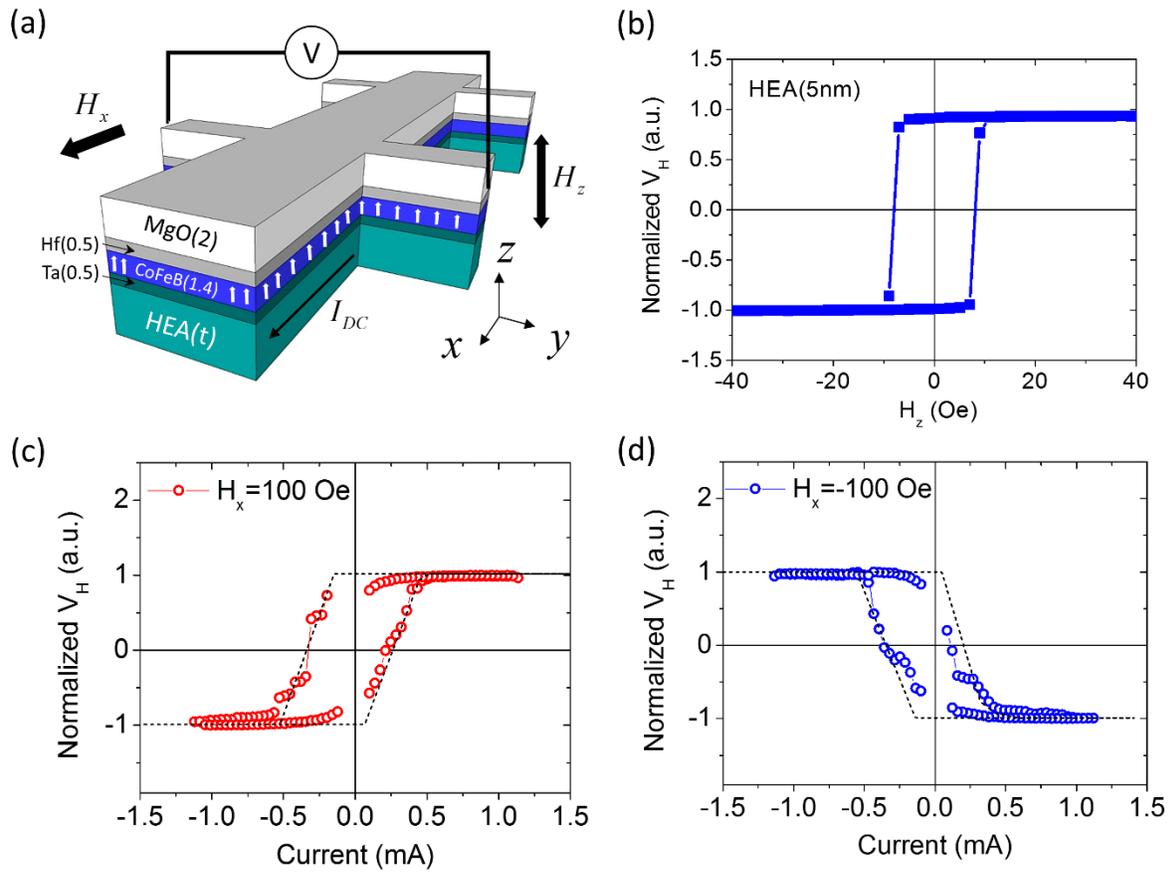

Figure 3. (a) Schematic illustration of an HEA($t_{HEA}$)/Ta(0.5)/CoFeB(1.4)/Hf(0.5)/MgO(2) Hall-bar device (Ta capping layer not shown). (b) A representative anomalous Hall voltage hysteresis loop obtained from a $t_{HEA} = 5\,\text{nm}$ Hall-bar device with PMA. Current-induced SOT-driven magnetization switching of a $t_{HEA} = 5\,\text{nm}$ Hall-bar device under in-plane bias field (c) $H_x = 100\,\text{Oe}$ and (d) $H_x = -100\,\text{Oe}$. The dashed lines serve as guide to the eye.

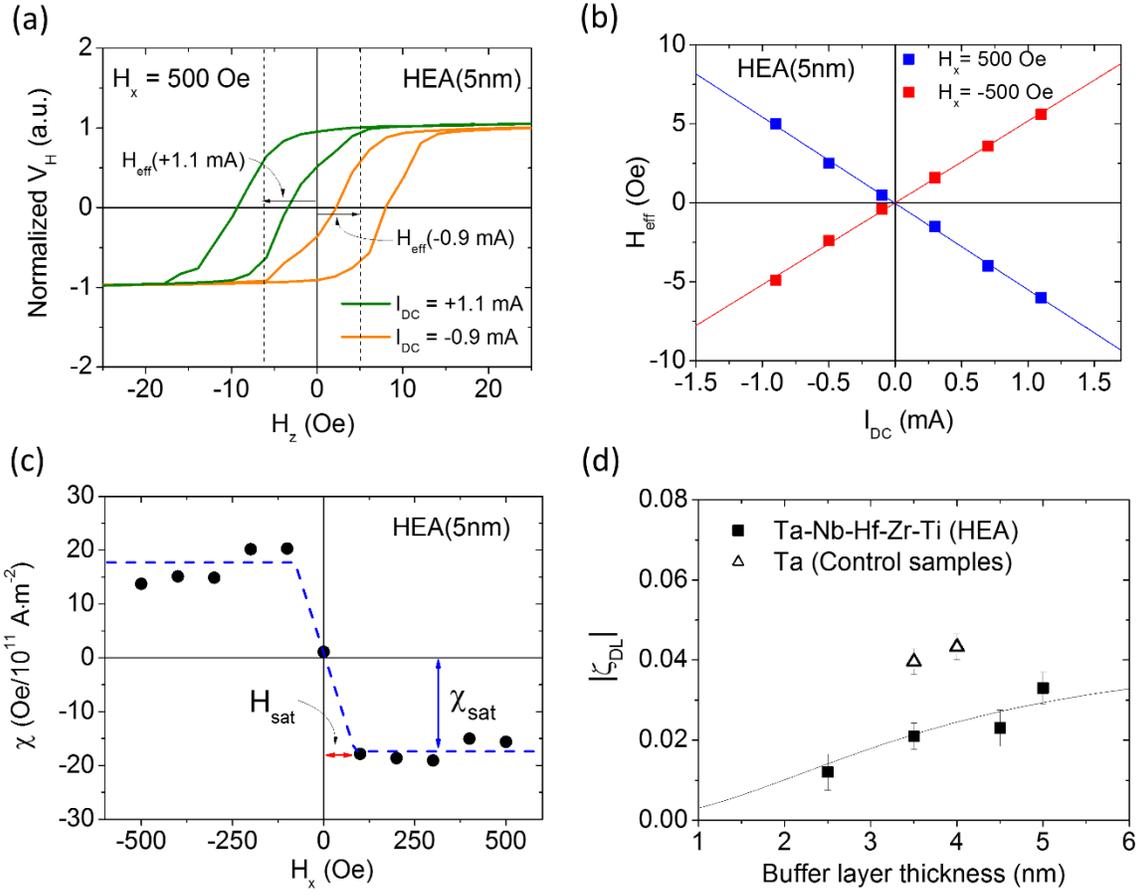

Figure 4. (a) Hysteresis loop shift measurement of the $t_{HEA} = 5\,\text{nm}$ HEA sample under in-plane bias field $H_x = 500\,\text{Oe}$ and applied DC current $I_{DC}$ with opposite polarities. (b) Hysteresis loop shift $H_{eff}$ as a function of applied current $I_{DC}$ with in-plane field $H_x = \pm 500\,\text{Oe}$. (c) Current-induced effective field per current density $\chi$ as a function of in-plane field. (d) The damping-like torque efficiencies $\zeta_{DL}$ for HEA (solid squares) and Ta control samples (open triangles) as functions of bottom layer thickness. The solid line represents fit to a spin diffusion model with spin diffusion length $\lambda_s^{HEA} = 2.5\,\text{nm}$ and $|\zeta_{DL}(t_{HEA} \to \infty)| = 0.04$.